\documentclass[twocolumn,showpacs,preprintnumbers,prd,nofootinbib]{revtex4}
\usepackage{epsfig,amsmath,amssymb,psfrag,pstricks}

\newcommand{\beq}{\begin{equation}}
\newcommand{\eeq}{\end{equation}}
\newcommand{\bea}{\begin{eqnarray}}
\newcommand{\eea}{\end{eqnarray}}
\newcommand{\Fig}[1]{Fig.\,\ref{#1}}

\newcommand{\Eq}[1]{Eq.\,(\ref{#1})}

\newcommand{\Tab}[1]{Tab.\,\ref{#1}}

\newcommand{\f}{\frac}

\newcommand{\gs}{g}
\newcommand{\as}{\alpha_s}

\newcommand{\MW}{M_{\scriptscriptstyle W}}

\newcommand{\mc}{m_c}
\newcommand{\mb}{m_b}

\newcommand{\muw}{\mu_{\scriptscriptstyle W}}

\newcommand{\muc}{\mu_c}
\newcommand{\mub}{\mu_b}

\newcommand{\Leff}{{\cal L}_{\rm eff}}

\newcommand{\GF}{G_F}

\newcommand{\GeV}{{\rm GeV}}
\newcommand{\TeV}{{\rm TeV}}

\newcommand{\ord}{{\cal O}}
\def\unit{\leavevmode\hbox{\small1\kern-3.6pt\normalsize1}}

\newcommand{\sL}{{\scalebox{0.6}{$L$}}}
\newcommand{\sR}{{\scalebox{0.6}{$R$}}}

\newcommand{\BXdga}{\bar{B} \to X_d \gamma}
\newcommand{\BXsga}{\bar{B} \to X_s \gamma}

\newcommand{\BXsll}{\bar{B} \to X_s l^+ l^-}

\newcommand{\Bsmm}{B_s \to \mu^+ \mu^-}

\newcommand{\BRga}{{\cal B} (\BXsga)}

\newcommand{\btosgamma}{b \to s \gamma}
\newcommand{\btosgluon}{b \to s g}

\newcommand{\Ztobb}{Z \to b \bar{b}}

\newcommand{\mysigma}{\hspace{0.4mm} \sigma}

\newcommand{\etal}{{\it et al}.}

\begin{document}

\allowdisplaybreaks

\preprint{ZU-TH 9/07} 

\title{Recent Developments in \boldmath $\BXsga$\footnote[2]{Based on
    invited talks given at the third meeting of the workshop ``Flavour
    in the era of the LHC'', CERN, Geneva, March 15--17, 2006 and at
    XLIInd Rencontres de Moriond, QCD and Hadronic Interactions, La
    Thuile, Italy, March 17--24, 2007.}  }

\author{Ulrich~Haisch}  

\affiliation{
Institut f\"ur Theoretische Physik, Universit\"at Z\"urich,
CH-8057 Z\"urich, Switzerland 
}

\date{\today}

\begin{abstract}
\noindent
We present a concise review of the recent theoretical progress
concerning the standard model calculation of the inclusive radiative
$\BXsga$ decay. Particular attention is thereby devoted to the
calculations of the next-to-next-to-leading order fixed-order $\ord
(\as^2)$ contributions, non-local $\ord (\as \Lambda/\mb)$ power
corrections, and logarithmic-enhanced $\ord (\as^2)$ cut-effects to
the decay rate. The current status of new physics calculations of the
inclusive $\btosgamma$ mode is also briefly summarized.
\end{abstract}


\maketitle

\section{ Introduction}
\label{sec:introduction}

As a flavor-changing-neutral-current process the inclusive radiative
$\bar{B}$-meson ($\bar{B} = \bar{B}^0$ or $B^-$) decay is
Cabibbo-Kobayashi-Maskawa- (CKM) and loop-suppressed within the
standard model (SM) and thus very sensitive to new physics (NP)
effects. In order to exploit the full potential of $\BXsga$ in
constraining the parameter space of beyond the SM physics both the
measurements and the SM prediction should be known as precisely as
possible.

The present experimental world average (WA) which includes the latest
measurements by CLEO, Belle, and BaBar \cite{CBB} is performed by the
Heavy Flavor Averaging Group \cite{Barbiero:2007cr} and reads for a
photon energy cut of $E_\gamma > E_{\rm cut}$ with $E_{\rm cut} = 1.6
\, \GeV$ in the $\bar{B}$-meson rest-frame
\bea \label{eq:WA} 
\BRga_{\rm exp} = \left ( 3.55 \pm 0.24^{+0.09}_{-0.10} \pm 0.03
\right ) \times 10^{-4} \, . \hspace{0.2mm} 
\eea
The total error of the WA is below $8 \%$ and consists of $i)$ a
combined statistical and systematic error, $ii)$ a systematic
uncertainty due to the extrapolation from $E_{\rm cut} = [1.8, 2.0] \,
\GeV$ to the reference value, and $iii)$ a systematic error due to the
subtraction of the $\BXdga$ event fraction. At the end of the
$B$-factory era the final accuracy of the averaged experimental value
is expected to be around $5 \%$. 

\section{Basic Properties of \boldmath $\BXsga$}
\label{sec:basics}

The $\btosgamma$ transition is dominated by perturbative QCD effects
which replace the power-like Glashow-Iliopoulos-Maiani (GIM)
suppression present in the electroweak (EW) vertex by a logarithmic
one. This mild suppression of the QCD corrected amplitude reduces the
sensitivity of the process to high scale physics, but enhances the
$\BXsga$ branching ratio (BR) with respect to the purely EW prediction
by a factor of around three. The logarithmic GIM cancellation
originates from the non-conservation of the tensor current which is
generated at the EW scale by loop diagrams involving $W$-boson and top
quarks exchange. The associated large logarithms $L = \ln \MW/\mb$
have to be resummed at each order in $\as$, using techniques of the
renormalization group (RG) improved perturbation theory. Factoring out
the Fermi constant $\GF$, the $\btosgamma$ amplitude receives
corrections of $\ord (\as^n L^n)$ at leading order (LO), of $\ord
(\as^n L^{n - 1})$ at next-to-leading order (NLO), and of $\ord (\as^n
L^{n - 2})$ at next-to-next-to-leading order (NNLO) in QCD. 

A suitable framework to achieve the necessary resummation is the
construction of an effective theory with five active quarks, photons
and gluons by integrating out the top quark and the EW
bosons. Including terms of dimension up to six in the local operator
product expansion (OPE) the relevant effective Lagrangian at a scale
$\mu$ reads
\beq \label{eq:Leff}
\Leff = {\cal L}_{{\rm QCD} \times {\rm QED}} + \f{4 \GF}{\sqrt{2}}
V_{ts}^\ast V_{tb} \sum_{k = 1}^8 C_k (\mu) Q_k \, .
\eeq
Here the first term is the conventional QCD and QED Lagrangian for the
light SM particles. In the second term $V_{ij}$ denotes the elements
of the CKM matrix and $C_k (\mu)$ are the Wilson coefficients of the
corresponding operators $Q_k$ built out of the light fields.

The operators and the numerical values of their Wilson coefficients at
$\mub \sim \mb$ are given by
\beq \label{eq:operators}
\begin{array}{l@{\hspace{5mm}}l} Q_{1,2} = (\bar{s} \Gamma_i c)
  (\bar{c} \Gamma^\prime_i b) \, , & C_{1,2} (\mb) \sim 1 \, , \\[1.5mm]
  Q_{3 \text{--} 6} = (\bar{s} \Gamma_i b) \sum_q (\bar{q}
  \Gamma^\prime_i q) \, , &
  \left | C_{3 \text{--} 6} (\mb) \right | < 0.07 \, , \\[1.5mm]
  Q_7 = \f{e \mb}{16 \pi^2} \bar{s}_{\sL} \sigma^{\mu \nu} b_{\sR}
  F_{\mu \nu} \, , &
  C_7 (\mb) \sim -0.3 \, , \\[1.5mm]
  Q_8 = \f{\gs \mb}{16 \pi^2} \bar{s}_{\sL} \sigma^{\mu \nu} T^a
  b_{\sR} G^a_{\mu \nu} \, , & C_8 (\mb) \sim -0.15 \, ,
\end{array} 
\eeq 
where $\Gamma$ and $\Gamma^\prime$, entering both the current-current
operators $Q_{1,2}$ and the QCD penguin operators $Q_{3 \text{--} 6}$,
stand for various products of Dirac and color matrices
\cite{Buras:2002tp}. In the dipole operator $Q_{7}$ $(Q_8)$, $e$
$(\gs)$ is the electromagnetic (strong) coupling constant, $q_{\sL,
  \sR}$ are the chiral quark fields, $F_{\mu \nu}$ $(G_{\mu \nu}^a)$
is the electromagnetic (gluonic) field strength tensor, and $T^a$ are
the color generators.

After including LO QCD effects the dominant contribution to the
partonic decay rate stems from charm quark loops that amount to $\sim
158 \%$ of the total $b \to s \gamma$ decay amplitude. The top
contribution is compared to the one from the charm quark with $\sim
-60 \%$ less than half as big and has the opposite sign. Diagrams
involving up quarks are suppressed by small CKM factors and lead at
the amplitude level to an effect of a mere $\sim 2\%$.

All perturbative calculations of $\btosgamma$ involve three steps:
$i)$ evaluation of the initial conditions $C_k (\muw)$ of the Wilson
coefficients at the matching scale $\muw \sim \MW$ by requiring
equality of Green's functions in the full and the effective theory up
to leading order in (external momenta)/$\MW$, $ii)$ calculation of the
anomalous dimension matrix (ADM) that determines the mixing and RG
evolution of $C_k (\mu)$ from $\muw$ down to the $\bar{B}$-meson scale
$\mub \sim \mb$, and $iii)$ determination of the on-shell matrix
elements of the various operators at $\mub \sim \mb$. Due to the
inclusive character of the $\BXsga$ mode and the heaviness of the
bottom quark, $\mb \gg \Lambda \sim \Lambda_{\rm QCD}$,
non-perturbative effects arise in the last step only as small
corrections to the partonic decay rate.

\section{\label{sec:progress} Theoretical Progress in \boldmath
  $\BXsga$}

At the NNLO level, the dipole and the four-quark operator matching
involves three and two loops, respectively. Renormalization constants
up to four loops must be found for $\btosgamma$ and $\btosgluon$
diagrams with four-quark operator insertions, while three-loop mixing
is sufficient in the remaining cases. Two-loop matrix elements of the
dipole and three-loop matrix elements of the four-quark operators must
be evaluated in the last step.

The necessary two- and three-loop matching was performed in
\cite{Bobeth:1999mk} and \cite{Misiak:2004ew}. The mixing at three
loops was determined in \cite{3mix} and at four loops in
\cite{Czakon:2006ss}. The two-loop matrix element of the photonic
dipole operator together with the corresponding bremsstrahlung was
found in \cite{1stQ7} and subsequently confirmed in
\cite{2ndQ7}. These calculations have been very recently extended to
include the full charm quark mass dependence
\cite{Asatrian:2006rq}. The three-loop matrix elements of the
current-current operators were derived in \cite{Bieri:2003ue} within
the so-called large-$\beta_0$ approximation. A calculation that goes
beyond this approximation employs an interpolation in the charm quark
mass \cite{Misiak:2006ab}. The effect of still unknown NNLO
contributions is believed to be smaller than the uncertainty that has
been estimated after incorporating the above corrections into the SM
calculation \cite{Misiak:2006ab, Misiak:2006zs}. To dispel possible
doubts about the correctness of this assumption, calculations of the
missing pieces are being pursued.

The most impressive bit of the various NNLO calculations is the one of
the four-loop ADM that describes the $\ord (\as^3)$ mixing of the
four-quark into the dipole operators \cite{Czakon:2006ss}. It has
involved the computation of more than 20000 four-loop diagrams and
required a mere computing time of several months on around 100 CPU's.

Another crucial part of the NNLO calculation is the interpolation in
the charm quark mass performed in \cite{Misiak:2006ab}. The three-loop
$\ord (\as^2)$ matrix elements of the current-current operators
contain the charm quark, and the NNLO calculation of these matrix
elements is essential to reduce the overall theoretical uncertainty of
the SM calculation. In fact, the largest part of the theoretical
uncertainty in the NLO analysis of the BR is related to the definition
of the mass of the charm quark \cite{Gambino:2001ew} that enters the
$\ord (\as)$ matrix elements $\langle s \gamma | Q_{1,2} | b
\rangle$. The latter matrix elements are non-vanishing at two loops
only and the scale at which $\mc$ should be normalized is therefore
undetermined at NLO. Since varying $\mc$ between $\mc (\mc) \sim 1.25
\, \GeV$ and $\mc (\mb) \sim 0.85 \, \GeV$ leads to a shift in the NLO
BR of more than $10 \%$ this issue is not an academic one. 

\begin{figure}
\begin{center}
\vspace{2mm}
\hspace{2mm} 
\mbox{
\includegraphics[height=1.65in,width=3in]{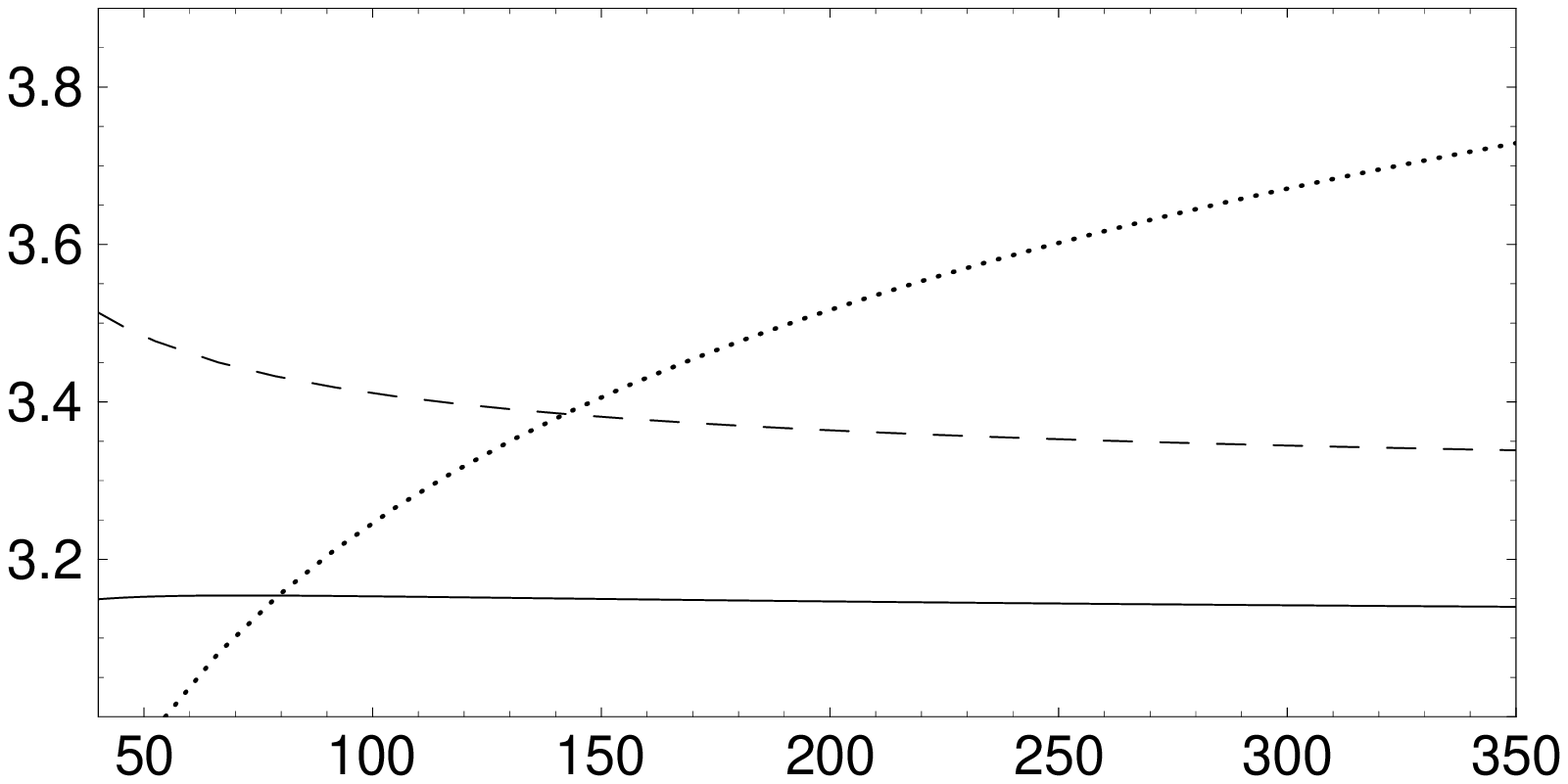}
} 

\vspace{1mm}
\hspace{2mm} 
\mbox{
\includegraphics[height=1.65in,width=3in]{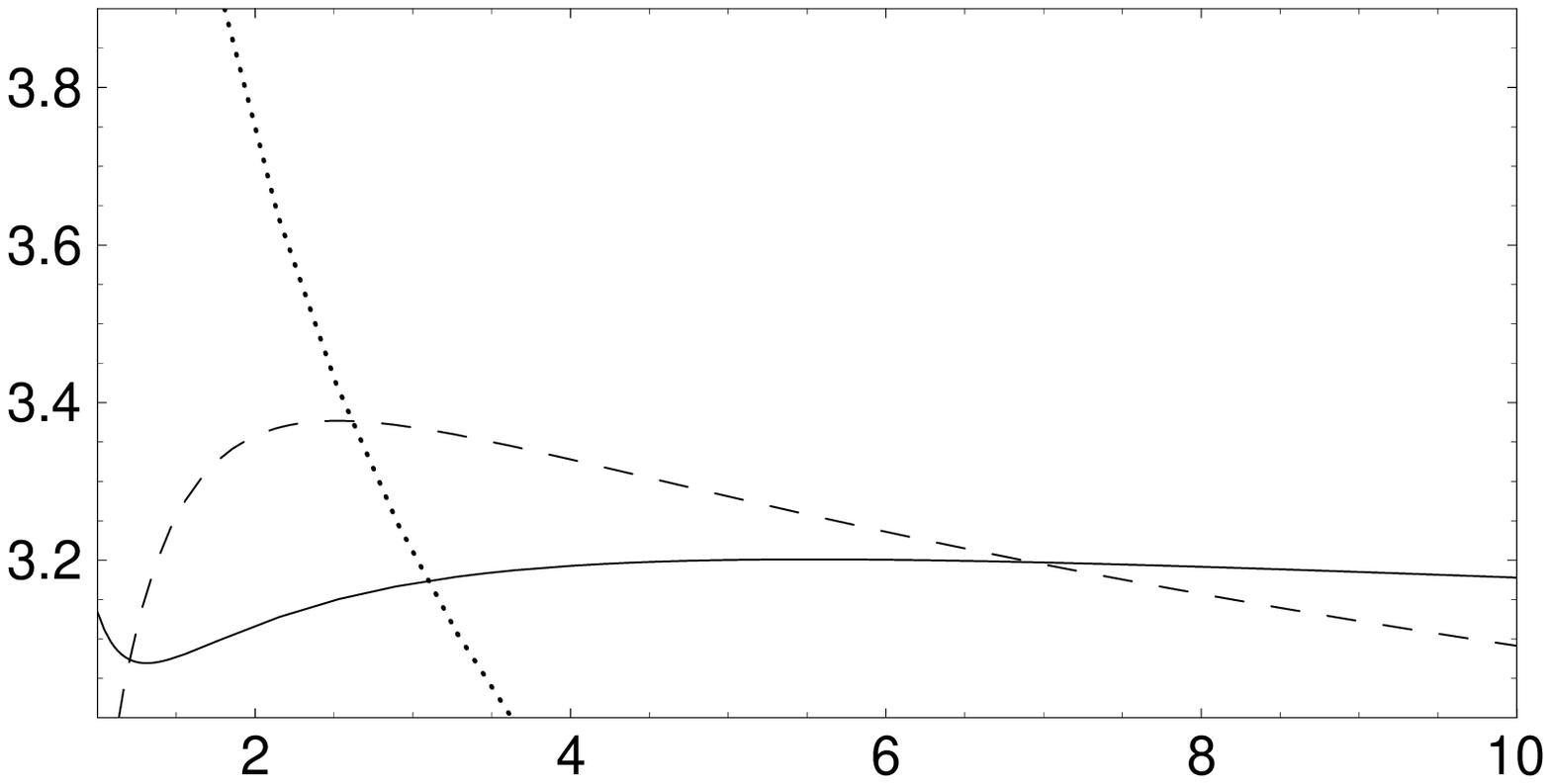}
} 

\vspace{1mm}
\hspace{2mm} 
\mbox{
\includegraphics[height=1.65in,width=3in]{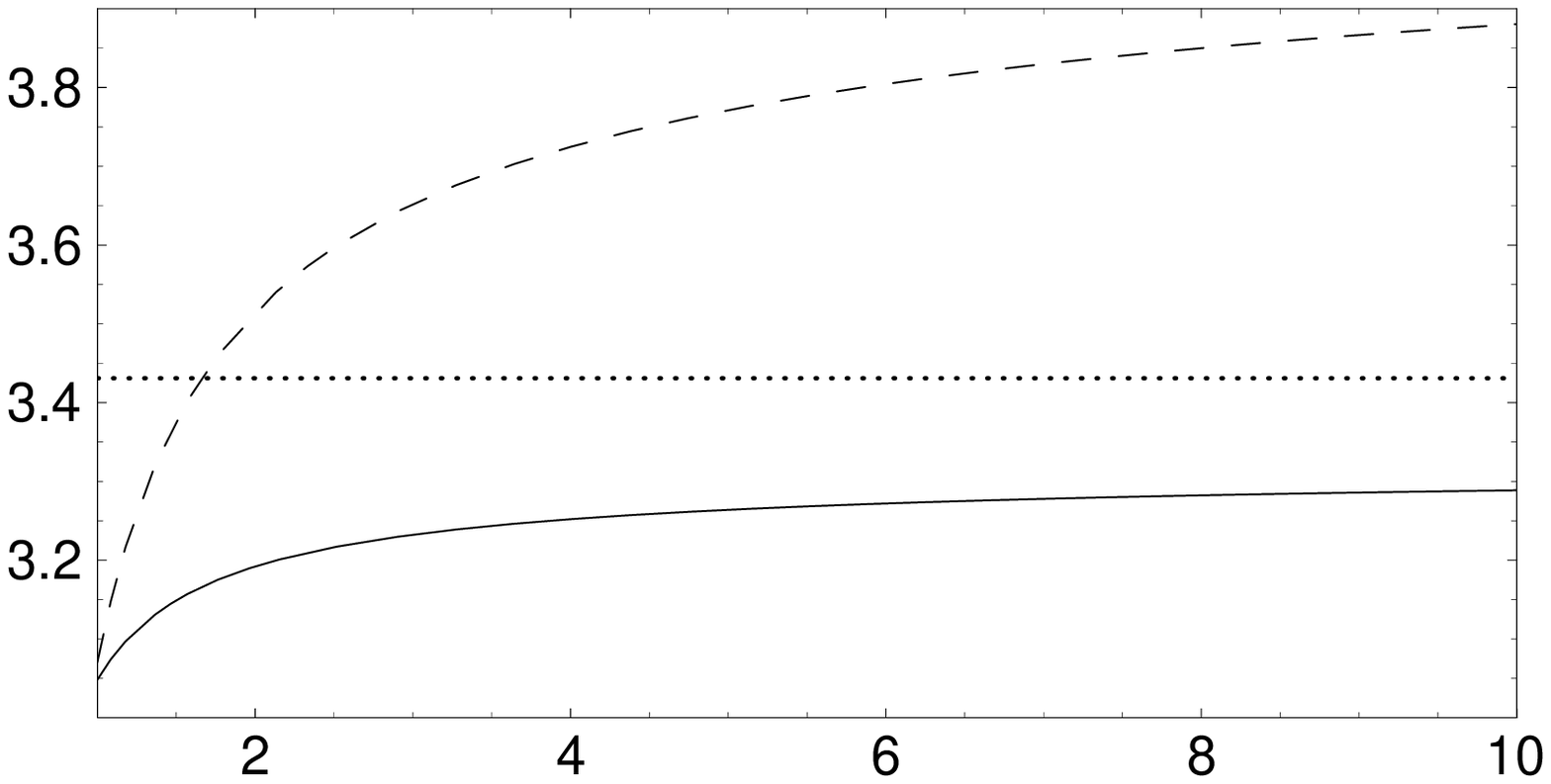}
}
\begin{pspicture}
\put(0,0){$\mu_{{\scriptscriptstyle W}, b, c} \ [\GeV]$}
\put(-4.95,1.00){\rotatebox{90}{$\BRga_{\rm SM} \ [10^{-4}]$}}
\put(-4.95,5.325){\rotatebox{90}{$\BRga_{\rm SM} \ [10^{-4}]$}}
\put(-4.95,9.625){\rotatebox{90}{$\BRga_{\rm SM} \ [10^{-4}]$}}
\end{pspicture}
\end{center}
\vspace{-4mm}
\caption{\sf Renormalization scale dependences of $\BRga_{\rm SM}$ at
  LO (dotted lines), NLO (dashed lines), and NNLO (solid lines) in
  QCD. The plots show from top to bottom the dependence on the
  matching scale $\muw$, the $\bar{B}$-meson scale $\mub$, and the
  charm quark mass renormalization scale $\muc$.}
\label{fig:scales}
\end{figure}
 
Finding the complete NNLO correction to $\langle s \gamma | Q_{1,2} |
b \rangle$ is a formidable task, since it involves the evaluation of
hundreds of three-loop on-shell vertex diagrams that are presently not
even known in the case $\mc = 0$. The approximation made in
\cite{Misiak:2006ab} is based on the observation that at the physical
point $\mc \sim 0.25 \, \mb$ the large $\mc \gg \mb$ asymptotic form
of the exact $\ord (\as)$ \cite{Buras:2002tp} and large-$\beta_0$
$\ord (\as^2 \beta_0)$ \cite{Bieri:2003ue} result matches the small
$\mc \ll \mb$ expansion rather well. This feature prompted the
analytic calculation of the leading term in the $\mc \gg \mb$
expansion of the three-loop diagrams, and to use the obtained
information to perform a interpolation to smaller values of $\mc$
assuming the $\ord (\as^2 \beta_0)$ part to be a good approximation of
the full $\ord(\as^2)$ result for vanishing charm quark mass. The
uncertainty related to this procedure has been assessed in
\cite{Misiak:2006ab} by employing three ans\"atze with different
boundary conditions at $\mc = 0$. A complete calculation of the $\ord
(\as^2)$ corrections to $\langle s \gamma | Q_{1,2} | b \rangle$ in
the latter limit or, if possible, for $\mc \sim 0.25 \, \mb$, would
resolve this ambiguity and should therefore be attempted.

Combining the aforementioned results it was possible to obtain the
first theoretical estimate of the total BR of $\BXsga$ at NNLO. For
the reference value $E_{\rm cut} = 1.6 \, \GeV$ the result of the
improved SM evaluation is given by \cite{Misiak:2006ab, Misiak:2006zs}
\beq \label{eq:NNLO}
\BRga_{\rm SM} = (3.15 \pm 0.23) \times 10^{-4} \, , 
\eeq
where the uncertainties from hadronic power corrections ($5 \%$),
parametric dependences ($3 \%$), higher-order perturbative effects $(3
\%)$, and the interpolation in the charm quark mass ($3 \%$) have been
added in quadrature to obtain the total error.

The reduction of the renormalization scale dependences at NNLO is
clearly seen in \Fig{fig:scales}. The most pronounced effect occurs in
the case of the charm quark mass renormalization scale $\muc$ that was
the main source of uncertainty at NLO. The current uncertainty of $3
\%$ due to higher-order effects is estimated from the variation of the
NNLO curves. The central value in \Eq{eq:NNLO} corresponds to the
choice $\mu_{{\scriptscriptstyle W}, b, c} = (160, 2.5, 1.5) \,
\GeV$. More details on the phenomenological analysis including the
list of input parameters can be found in \cite{Misiak:2006ab}.

It is well-known that the OPE for $\BXsga$ has certain limitations
which stem from the fact that the photon has a partonic
substructure. In particular, the local expansion does not apply to
contributions from operators other than $Q_7$, in which the photon
couples to light quarks \cite{Kapustin:1995fk, lightquarks}. While the
presence of non-local power corrections was thus foreseen such terms
have been studied until recently only in the case of the $(Q_8, Q_8)$
interference \cite{Kapustin:1995fk}. In \cite{Lee:2006wn} the analysis
of non-perturbative effects that go beyond the local OPE have been
extended to the enhanced non-local terms emerging from $(Q_7, Q_8)$
insertions. The found correction scales like $\ord (\as
\Lambda/\mb)$ and its effect on the BR was estimated using the vacuum
insertion approximation to be $-[0.3, 3.0] \%$. A measurement of the
flavor asymmetry between $\bar{B}^0 \to X_s \gamma$ and $B^- \to X_s
\gamma$ could help to sustain this numerical estimate
\cite{Lee:2006wn}. Potentially as or maybe even more important than
the latter correction are those arising from the $(Q_{1,2}, Q_7)$
interference. Naive dimensional analysis suggests that some
non-perturbative corrections to them also scale like $\ord (\as
\Lambda/\mb)$. Since at the moment there is not even an estimate
of those corrections, a non-perturbative uncertainty of $5 \%$ has
been assigned to the result in \Eq{eq:NNLO}. This error is the
dominant theoretical uncertainty at present and thought to include all
known \cite{Lee:2006wn} and unknown $\ord (\as \Lambda/\mb)$
terms. Calculating the precise impact of the enhanced non-local power
corrections may remain notoriously difficult given the limited control
over non-perturbative effects on the light cone.

A further complication in the calculation of $\BXsga$ arises from the
fact that all measurements impose stringent cuts on the photon energy
to suppress the background from other $\bar{B}$-meson decay
processes. Restricting $E_\gamma$ to be close to the physical endpoint
$E_{\rm max} = m_B/2$, leads to a breakdown of the local OPE, which
can be cured by resummation of an infinite set of leading-twist terms
into a non-perturbative shape function \cite{shapefunction}. A
detailed knowledge of the shape function and other subleading effects
is required to extrapolate the measurements to a region where the
conventional OPE can be trusted.

The transition from the shape function to the OPE region can be
described by a multi-scale OPE (MSOPE) \cite{Neubert:2004dd}. In
addition to the hard scale $\mu_h \sim \mb \sim 5 \, \GeV$, this
expansion involves a hard-collinear scale $\mu_{hc} \sim \sqrt{\mb
  \Delta} \sim 2.5 \, \GeV$ corresponding to the typical hadronic
invariant mass of the final state $X_s$, and a soft scale $\mu_s \sim
\Delta \sim 1.5 \, \GeV$ related to the width $\Delta/2 = \mb/2 -
E_{\rm cut}$ of the energy window in which the photon spectrum is
measured. In the MSOPE framework, the perturbative tail of the
spectrum receives calculable corrections at all three scales, and may
be subject to large perturbative corrections due to the presence of
terms proportional to $\as (\sqrt{\mb \Delta}) \sim 0.27$ and $\as
(\Delta) \sim 0.36$.

A systematic MSOPE analysis of the $(Q_7, Q_7)$ interference at NNLO
has been performed in \cite{Becher:2006pu}. Besides the hard matching
corrections, it involves the two-loop logarithmic and constant terms
of the jet \cite{Neubert:2004dd, jet} and soft function \cite{soft}.
The three-loop ADM of the shape function remains unknown and is not
included. The MSOPE result can be combined with the fixed-order
prediction by computing the fraction of events $1 - T$ that lies in
the range $E_{\rm cut} = [1.0, 1.6] \, \GeV$. The analysis
\cite{Becher:2006pu} yields 
\beq \label{eq:T}
1 - T = 0.07{^{+0.03}_{-0.05}}_{\rm pert} \pm 0.02_{\rm hadr} \pm
0.02_{\rm pars} \, ,  
\eeq 
where the individual errors are perturbative, hadronic, and
parametric. The quoted value is almost twice as large as the NNLO
estimate $1 - T = 0.04 \pm 0.01_{\rm pert}$ obtained in fixed-order
perturbation theory \cite{Misiak:2006ab, Misiak:2006zs, Mikolaj} and
plagued by a significant additional theoretical error related to
low-scale perturbative corrections. These large residual scale
uncertainties indicate a slow convergence of the MSOPE series
expansion in the tail region of the photon energy spectrum. Given that
$\Delta$ is always larger than $1.4 \, \GeV$ and thus fully in the
perturbative regime this feature is unexpected.

Additional theoretical information on the shape of the photon energy
spectrum can be obtained from the universality of soft and collinear
gluon radiation. Such an approach can be used to predict large
logarithms of the form $\ln (E_{\rm max} - E_{\rm cut})$. These
computations have also achieved NNLO accuracy \cite{A&E} and
incorporate Sudakov and renormalon resummation via dressed gluon
exponentiation (DGE) \cite{A&E, Gardi:2006jc}. The present NNLO
estimate of $1 - T = 0.016 \pm 0.003_{\rm pert}$ \cite{A&E, Einan}
indicates a much thinner tail of the photon energy spectrum and a
considerable smaller perturbative uncertainty than reported in
\cite{Becher:2006pu}. The DGE analysis thus supports the view that the
integrated photon energy spectrum below $E_{\rm cut} = 1.6 \, \GeV$ is
well approximated by a fixed-order perturbative calculation,
complemented by local OPE power corrections. To understand how
precisely the tail of the photon energy spectrum can be calculated
requires nevertheless further theoretical investigations.

\begin{widetext}
\begin{center}
\begin{table}[!t]      
\begin{center}
\begin{tabular}{|@{\hspace{2.5mm}}c@{\hspace{2.5mm}}|@{\hspace{2.5mm}}c@{\hspace{2.5mm}}|@{\hspace{2.5mm}}c@{\hspace{2.5mm}}|@{\hspace{2.5mm}}c@{\hspace{2.5mm}}|}
  \hline \\[-4.5mm]
  Model & Accuracy & Effect & Bound \\[0.5mm] 
  \hline \hline 
  THDM type II & NLO \cite{THDM, Bobeth:1999ww} & $\Uparrow$ & 
  $M_H^\pm > 295 \, \GeV \ (95 \% \ {\rm CL})$ \cite{Misiak:2006zs} \\   
  \hline
  MFV MSSM & NLO \cite{Bobeth:1999ww, 
    Ciuchini:1998xy, Borzumati:2003rr, Degrassi:2006eh, LTB, 
    D'Ambrosio:2002ex} & $\Updownarrow$  
  & --- \\     
  \hline
  LR & NLO \cite{Bobeth:1999ww} & $\Updownarrow$ & --- \\  
  \hline
  general MSSM & LO \cite{GMSSM} & $\Updownarrow$ & 
  $\begin{array}{ll} | (\delta_{23}^d)_{LL} | \lesssim 4 \times 10^{-1}, 
    &  | (\delta_{23}^d)_{RR} | \lesssim 8 \times 10^{-1}, \\ 
    | (\delta_{23}^d)_{LR} | \lesssim 6 \times 10^{-2}, 
    &  | (\delta_{23}^d)_{RL} | \lesssim 2 \times 10^{-2} \end{array}$ 
  \cite{GMSSM} \\  
  \hline 
  mUED & LO \cite{mUED} & $\Downarrow$ & $1/R > 600 \, \GeV \ (95 \% \
  {\rm CL})$ \cite{Haisch:2007vb} \\  
  \hline  
  RS & LO \cite{RS} & $\Uparrow$ & $M_{\rm KK} \gtrsim 2.4 \, \TeV$ \\  
  \hline
  LH & LO \cite{LH} & $\uparrow$ & --- \\  
  \hline 
  LHT & LO \cite{Blanke:2006sb} & $\updownarrow$ & --- \\
  \hline 
\end{tabular}
\end{center}
\vspace{0mm}
\caption{\sf Theoretical accuracy, effect on $\BRga$ relative to the SM
  prediction, and if applicable, constraint on the parameter space
  following from $\BXsga$ in popular NP scenarios. Arrows pointing
  upward (downward) indicate that the NP effects interfere
  constructively (destructively) with the SM $\btosgamma$
  amplitude. Single (double) arrows specify whether the maximal
  possible shift is smaller (larger) than the theoretical
  uncertainty of the SM expectation. See text for details.}
\label{tab:NP}
\end{table}
\end{center}
\end{widetext}

\section{\label{sec:newphysics} New Physics in \boldmath $\BXsga$}

Compared with the experimental WA of \Eq{eq:WA}, the new SM prediction
of \Eq{eq:NNLO} is lower by more than $1 \mysigma$. Potential beyond
SM contributions should now be preferably constructive, while models
that lead to a suppression of the $\btosgamma$ amplitude are more
severely constrained than in the past, where the theoretical
determination used to be above the experimental one
\cite{Buras:2002tp}.

NP affects the initial conditions of the Wilson coefficients of the
operators in the low-energy effective theory and might also induce new
operators besides those already present in the SM. Complete NLO
matching calculations are available only in the case of the
two-Higgs-doublet models (THDMs) \cite{THDM, Bobeth:1999ww}, the
minimal supersymmetric SM (MSSM) with minimal-flavor-violation (MFV)
for small and large $\tan \beta$ \cite{Bobeth:1999ww, Ciuchini:1998xy,
  Borzumati:2003rr, Degrassi:2006eh, LTB, D'Ambrosio:2002ex}, and
left-right (LR) symmetric models \cite{Bobeth:1999ww}. In the general
MSSM \cite{GMSSM}, extra dimensional models like minimal universal
extra dimensions (mUED) \cite{mUED} or Randall-Sundrum (RS) scenarios
\cite{RS}, and littlest Higgs (LH) models without \cite{LH} and with
$T$-parity (LHT) \cite{Blanke:2006sb}, the accuracy is in general
strictly LO and hence far from the one achieved in the SM. The main
features and results of recent analyses of beyond SM physics in
$\BXsga$ are listed in \Tab{tab:NP}. In the following we will briefly
review the most important findings.

\begin{figure}
\begin{center}
\vspace{-7mm}
\makebox{
\makebox{\hspace{4mm} \includegraphics[width=2.85in,height=2.75in]{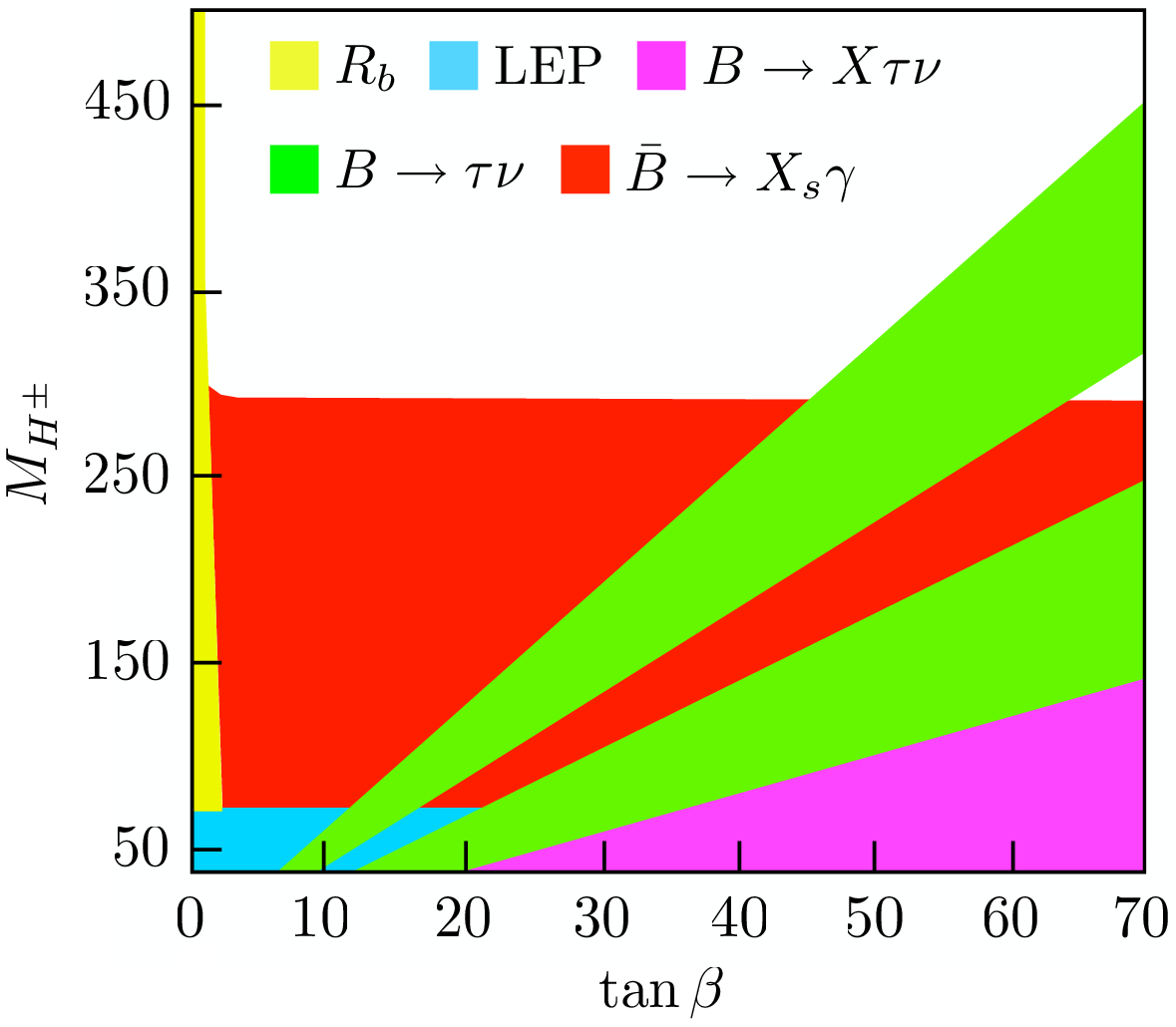}} 
} 
\makebox{
\begin{psfrags}
\providecommand{\psfragtextscale}{1.2}
\providecommand{\psfragmathscale}{\psfragtextscale}
\providecommand{\psfragnumericscale}{\psfragtextscale}
\providecommand{\psfragtextstyle}{}
\providecommand{\psfragmathstyle}{}
\providecommand{\psfragnumericstyle}{}

\psfrag{x}[cc][cc][0.8][0]{${\cal B} (\bar{B} \to X_s \gamma)_{\rm SM}~[10^{-4}]
$}
\psfrag{y}[bc][bc][0.8][0]{$\Delta {\cal B} (\bar{B} \to X_s \gamma)_{\rm SM}~[1
0^{-4}]$}
\makebox{\hspace{-1cm} \includegraphics[width=2.5in,height=2.5in]{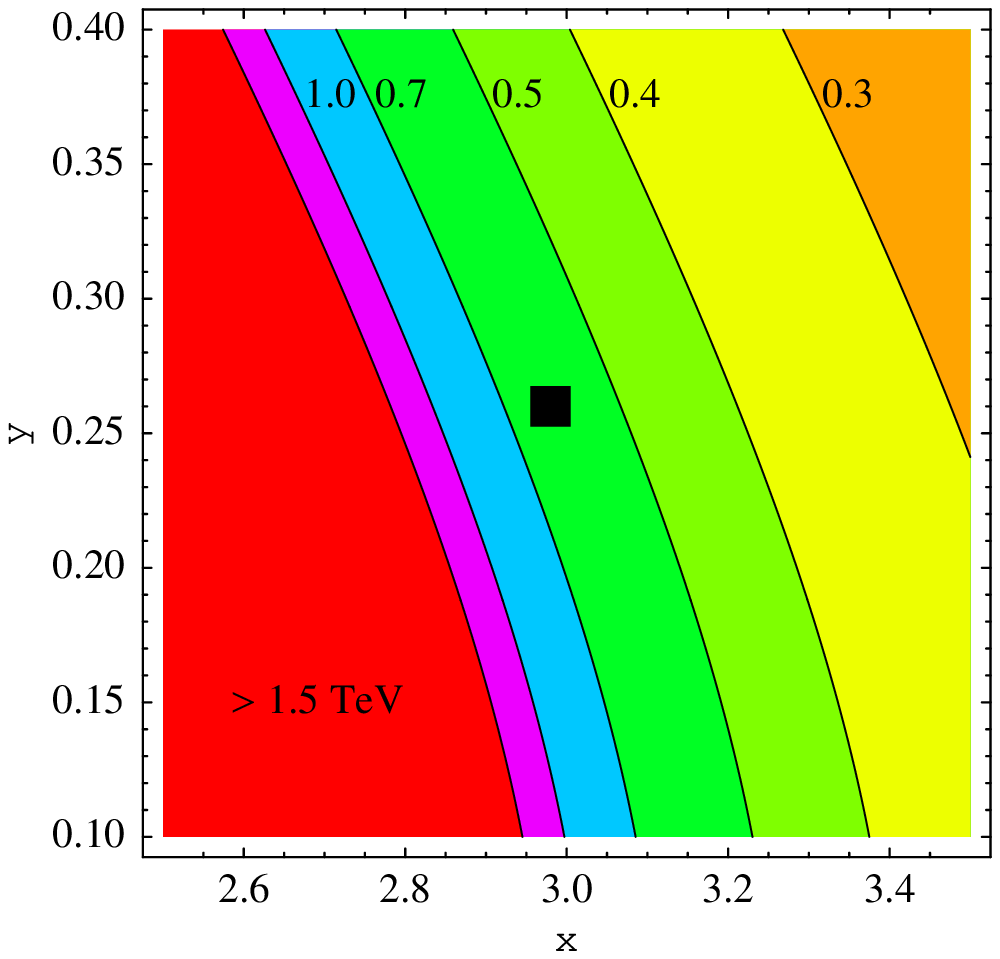}} 
\end{psfrags}
}
\end{center}
\vspace{-6mm}
\caption{\sf Top: Direct and indirect bounds on $M_{H^\pm}$ in the
  THDM type II model as a function of $\tan \beta$. The colored areas
  are excluded by the constraints at $95 \%$ CL. Bottom: $95 \%$ CL
  limits on the compactification scale $1/R$ in the mUED model as a
  function of the SM central value and total error. The present SM
  result is indicated by the black square.  See text for details.}
\label{fig:NPbounds}
\end{figure}

Even though the effect of charged Higgs boson contributions in the
THDM type II model is necessarily constructive \cite{THDM,
  Bobeth:1999ww}, the lower bound on $M_{H^\pm}$ following from
$\BXsga$ remains in general stronger than all other direct and
indirect constraints. In particular, $\BXsga$ still prevails over $B
\to \tau \nu$ \cite{Btaunu} for all values of $\tan \beta$ apart from
those lying in the range $\tan \beta \sim [45, 65]$. This is
illustrated in the upper panel of \Fig{fig:NPbounds}. The derived $95
\%$ confidence level (CL) limit amounts to $M_{H^\pm} > 295 \, \GeV$
independently of $\tan \beta$ \cite{Misiak:2006zs}. In the THDM type I
model, the strongest constraint on $M_{H^\pm}$ stems from the ratio of
the widths of the $Z$-boson decay into bottom quarks and hadrons,
$R_b$, and not from $\BXsga$.

In the MFV MSSM the complete NLO corrections to $\BXsga$ are also
known. The needed two-loop diagrams containing gluons and gluinos were
evaluated in \cite{Bobeth:1999ww, Ciuchini:1998xy} and
\cite{Borzumati:2003rr, Degrassi:2006eh}, respectively. Since EW
interactions affect the quark and squark mass matrices in a different
way, their alignment is not RG invariant and MFV can only be imposed
at a certain scale $\mu_{\rm MFV}$ that is related to the mechanism of
supersymmetry (SUSY) breaking \cite{Degrassi:2006eh}. For $\mu_{\rm
  MFV}$ much larger than the SUSY masses $M_{\rm SUSY}$, the ensuing
large logarithms can lead to sizable effects in $\BXsga$, and need to
be resummed by solving the RG equation of the flavor-changing
gluino-quark-squark couplings.

In the limit of $M_{\rm SUSY} \gg \MW$, SUSY effects can be absorbed
into the coupling constants of local operators in an effective theory
\cite{LTB, D'Ambrosio:2002ex}. The Higgs sector of the MSSM is
modified by these non-decoupling corrections and can differ notably
from the native THDM type II model. Some of the corrections to
$\BXsga$ in the effective theory are enhanced by $\tan \beta$. As a
result, they can be sizable, of order $\as \tan \beta \sim 1$ for
values of $\tan \beta \gg 1$, and need to be resummed if applicable.
In the large $\tan \beta$ regime the relative sign of the chargino
contribution is given by $-{\rm sgn}(A_t \mu)$. For ${\rm sgn} (A_t
\mu) > 0$, the chargino and charged Higgs contributions interfere
hence constructively with the SM result and this tends to rule out
large positive values of the product of the trilinear soft SUSY
breaking coupling $A_t$ and the Higgsino parameter $\mu$.

In the MSSM with generic sources of flavor violation a complete NLO
analysis is still missing up to date. Experimental constraints on
generic $b \to s$ flavor violation have been studied extensively
\cite{GMSSM}, and radiative inclusive $\bar{B}$-meson decays play a
central role in these analyses. In particular, for small and moderate
values of $\tan \beta$ all four mass insertions $(\delta_{23}^d)_{AB}$
with $A, B = L, R$ except for $(\delta_{23}^d)_{RR}$ are determined
entirely by $\BXsga$. The bounds on the mass insertions
$(\delta_{23}^d)_{AB}$ corresponding to $\tan \beta = 10$ are given in
\Tab{tab:NP}. For large values of $\tan \beta$ neutral Higgs penguin
contributions become important and the constraints from both $\Bsmm$
and $B_s \text{--} \bar{B}_s$ mixing surpass the one from
$\BXsga$. The effect of the precision measurement of the mass
difference $\Delta M_s$ \cite{Abulencia:2006ze} is especially strong
in the case of $(\delta_{23}^d)_{RL, RR}$. At large $\tan \beta$ the
limits on both mass insertions are now imposed by the $B_s \text{--}
\bar{B}_s$ mixing constraint alone.

Since Kaluza-Klein (KK) modes in the mUED model interfere
destructively with the SM $\btosgamma$ amplitude \cite{mUED}, $\BXsga$
leads to a very powerful bound on the inverse compactification radius
of $1/R > 600 \, \GeV$ at $95 \%$ CL \cite{Haisch:2007vb}. This
exclusion is independent from the Higgs mass and therefore stronger
than any limit that can be derived from EW precision measurements. The
$95 \%$ CL bound on $1/R$ as a function of the SM central value and
error is shown in the lower panel of \Fig{fig:NPbounds}. In RS models,
KK modes enhance the BR relative to the SM \cite{RS}, and the bound on
the KK masses is in consequence with $M_{\rm KK} \gtrsim 2.4 \, \TeV$
significantly weaker than the constraint that derives from EW
precision data.

The contributions to $\BXsga$ from new heavy vector bosons, scalars,
and quarks appearing in LH models, have been studied in \cite{LH} for
the original model, and in \cite{Blanke:2006sb} for an extension in
which an additional $Z_2$ symmetry called $T$-parity is introduced to
preserve custodial $SU (2)$ symmetry. While in the former case the new
contributions always lead to an enhancement of $\BRga$ \cite{LH}, in
the latter case also a suppression with respect to the SM expectation
is possible \cite{Blanke:2006sb}. As the found LH effects in $\BXsga$
are generically smaller than the theoretical uncertainties in the SM,
they essentially do not lead to any restriction on the parameter
space.

An alternative avenue to NP analyses of $\BXsga$ consists in
constraining the Wilson coefficients of the operators in the
low-energy effective theory. This model-independent approach has been
applied combining various $B$- and $K$-meson decay modes both
neglecting \cite{modelindependent, Haisch:2007ia} and including
\cite{D'Ambrosio:2002ex, Fox:2007in} operators that do not contribute
in the SM. In particular, in the former case, merging the information
on $\BXsga$ with the one on $\BXsll$ \cite{BXsll}, one can infer that
the sign of the $\btosgamma$ amplitude is in all probability SM-like
\cite{Gambino:2004mv}. In the case of the $Z$-penguin amplitude the
same conclusion can be drawn on the basis of the precision
measurements of $R_b$ and the other $\Ztobb$ pseudo observables
\cite{Haisch:2007ia}.
 
\section{\label{sec:conclusions} Conclusions}

The inclusion of NNLO QCD corrections has lead to a significant
suppression of the renormalization scale dependences of the $\BXsga$
branching ratio that have been the main source of theoretical
uncertainty at NLO. The central value of the SM prediction is shifted
downward relative to all previously published NLO results. It is now
more than $1 \mysigma$ below the experimental average. This revives
the possibility for explorations of new physics contributions to rare
flavor-changing $B$-decay processes. The dominant theoretical
uncertainty in the SM is currently due to unknown non-perturbative
effects. A reduction of this error, together with a calculation of the
three-loop matrix elements of the current-current operators and a
better understanding of the tail of the photon energy spectrum is
essential to further increase the power of $\BXsga$ in the search for
new physics. 

\section*{\label{sec:acknowledgements} Acknowledgements}

I am grateful to P.~Gambino and M.~Misiak for valuable comments on the
manuscript and useful discussions. Private communications with
E.~Gardi and M.~Misiak are acknowledged. This work has been supported
in part by the European Union and the Schweizer Nationalfonds.

\end{document}